\documentclass[twocolumn]{aastex631}

\DeclareUnicodeCharacter{2212}{\ensuremath{-}}

\usepackage{array}
\usepackage{amsmath}
\usepackage{hyperref}
\usepackage{longtable}
\usepackage{rotating, graphicx}
\newcolumntype{P}[1]{>{\centering\arraybackslash}p{#1}}
\newcolumntype{R}[1]{>{\raggedleft\arraybackslash}p{#1}}
\newcolumntype{L}[1]{>{\raggedright\arraybackslash}p{#1}}

\received{May 20, 2024}
\revised{July 15, 2024}
\accepted{July 16, 2024}
\submitjournal{ApJ}

\shorttitle{}
\shortauthors{Gibbs et al.}

\begin{document}

\title{Subaru/CHARIS High-Resolution Mode Spectroscopy of the Brown Dwarf Companion HD 33632 Ab}



\author[0000-0002-9027-4456]{Aidan B. Gibbs}
\affiliation{Department of Physics \& Astronomy, University of California, Los Angeles, CA 90095, USA}

\author[0000-0002-8984-4319]{Briley L. Lewis}
\affiliation{Department of Physics \& Astronomy, University of California, Los Angeles, CA 90095, USA}

\author[0000-0002-0176-8973]{Michael P. Fitzgerald}
\affiliation{Department of Physics \& Astronomy, University of California, Los Angeles, CA 90095, USA}

\author[0000-0003-2630-8073]{Timothy D. Brandt}
\affiliation{Department of Physics, University of California, Santa Barbara, Santa Barbara, CA 93106, USA}

\author[0000-0001-8892-4045]{Minghan Chen}
\affiliation{Department of Physics, University of California, Santa Barbara, Santa Barbara, CA 93106, USA}

\author[0000-0002-6845-9702]{Yiting Li}
\affiliation{Department of Astronomy, University of Michigan, Ann Arbor, MI 48109, USA}

\author[0000-0001-5831-9530]{Rachel Bowens-Rubin}
\affiliation{Department of Astronomy, University of California, Santa Cruz, Santa Cruz, CA 95064, USA}

\author[0000-0003-0054-2953]{Rebecca Jensen-Clem}
\affiliation{Department of Astronomy, University of California, Santa Cruz, Santa Cruz, CA 95064, USA}

\author[0000-0003-0526-1114]{Benjamin A. Mazin}
\affiliation{Department of Physics, University of California, Santa Barbara, Santa Barbara, CA 93106, USA}

\begin{abstract}

Brown dwarfs (BD) are model degenerate in age and mass. High-contrast imaging and spectroscopy of BD companions to host stars where the mass and age can be independently constrained by dynamics and stellar age indicators respectively provide valuable tests of BD evolution models. In this paper, we present a new epoch of Subaru/CHARIS \textit{H}- and \textit{K}-band observations of one such previously discovered system, HD 33632 Ab. We reanalyze the mass and orbit using our new epoch of extracted relative astrometry, and fit extracted spectra to the newest generation of equilibrium, disequilibrium, and cloudy spectral and evolution models for BDs. No spectral model perfectly agrees with evolutionary tracks and the derived mass and age, instead favoring a somewhat younger BD than the host star's inferred age. This tension can potentially be resolved using atmosphere and evolution models that consider both clouds and disequilibrium chemistry simultaneously, or by additional future spectra at higher resolution or in other band passes. Photometric measurements alone remain consistent with the luminosity predicted by evolutionary tracks. Our work highlights the importance of considering complexities like clouds, disequilibrium chemistry, and composition when comparing spectral models to evolutionary tracks.

\end{abstract}



\section{Introduction} \label{sec:intro}

\begin{deluxetable*}{llll}[t]\label{tab:props}
\centering
\tablewidth{\textwidth}
\tablecaption{HD 33632 System Properties}
\tablehead{\multicolumn{4}{c}{HD 33632 A}}
\startdata
Mass ($M_{\odot}$) & $1.1\pm0.1$ & \multicolumn{2}{r}{\citep{MBrandt2021}} \\
$\log_{10}(L/L_{\odot})$ & $-4.62^{+0.04}_{-0.08}$ & \multicolumn{2}{r}{\citep{Currie2020}} \\
Distance (pc) & $26.39\pm0.18$ & \multicolumn{2}{r}{\citep{Gaia2020_parallax}} \\
Age (Gyr) & $1.7\pm0.4$ & \multicolumn{2}{r}{\citep{MBrandt2021}} \\
\text{[Fe/H]} (dex) & $-0.22\pm0.03$ & \multicolumn{2}{r}{\citep{Soubiran2016_PASTEL}} \\
$J_\text{2MASS}$ & $5.430\pm0.020$ & \multicolumn{2}{r}{\citep{2MASS2003}} \\
$H_\text{2MASS}$ & $5.193\pm0.015$ & \multicolumn{2}{r}{\citep{2MASS2003}} \\
$K_\text{S,2MASS}$ & $5.166\pm0.020$ & \multicolumn{2}{r}{\citep{2MASS2003}} \\
\hline
\multicolumn{4}{c}{HD 33632 B} \\
\hline
Mass ($M_{\odot}$) & $0.215\pm0.029$ & \multicolumn{2}{r}{\citep{MBrandt2021}} \\
Separation ($''$) & $33.99086\pm0.00003$ & \multicolumn{2}{r}{\citep{Gaia2020_parallax}} \\
$J_\text{2MASS}$ & $10.38\pm0.03$ & \multicolumn{2}{r}{\citep{2MASS2003}} \\
$H_\text{2MASS}$ & $9.863\pm0.019$ & \multicolumn{2}{r}{\citep{2MASS2003}} \\
\hline
\multicolumn{4}{c}{HD 33632 Ab} \\
\hline
 & \multicolumn{1}{l}{\citep{Currie2020}} & \multicolumn{1}{l}{\citep{MBrandt2021}} & \multicolumn{1}{l}{This Work} \\
 \cline{2-4} 
 $J_\text{MKO}$ & $16.91\pm0.11$ & ... & ... \\
 $H_\text{MKO}$ & $16.00\pm0.09$ & ... & $15.96\pm0.08$ \\
 $K_\text{S,MKO}$ & $15.37\pm0.09$ & ... & $15.54\pm0.12$ \\
 $L_\text{P,MKO}$ & $13.67\pm0.15$ & ... & ... \\
 \cline{2-4} 
 & \multicolumn{3}{c}{Fitted Parameters} \\
  \cline{2-4} 
 Mass ($M_{J}$) & $46.4^{+8.1}_{-7.5}$ & $50.0^{+5.5}_{-5.0}$ & $51.6^{+5.4}_{-4.8}$ \\
 a (au) & $21.2^{+4.1}_{-4.5}$ & $23.6^{+3.2}_{-4.5}$ & $23.7^{+3.1}_{-3.8}$ \\
 e & $0.18^{+0.20}_{-0.14}$ & $0.12^{+0.18}_{-0.09}$ & $0.143^{+0.13}_{-0.089}$ \\
 i ($^\circ$) & $39.4^{+8.0}_{-20}$ & $45.2^{+4.7}_{-11}$ & $42.2^{+5.5}_{-11.0}$ \\
 $\omega$ ($^\circ$) & $151^{+155}_{-131}$ & $0^{+86}_{-140}$ & $282^{+60}_{-252}$ \\
 $T_0$ (JD) & $2440900^{+4500}_{-13100}$ & $2468815^{+18000}_{-5800}$ & $2466542^{+3096}_{-5100}$ \\
 PA ascending ($^{\circ}$) & $38.2^{+7.2}_{-7.0}$ & $39.3^{+5.7}_{-6.5}$ & $38.0^{+5.3}_{-5.3}$ \\
 $\lambda_\text{ref}$ ($^\circ$) & $21^{+14}_{-10}$ & $202^{+14}_{-9.5}$ & $208^{+10}_{-8}$ \\
 Period (yrs) & $91^{+27}_{-27}$ & $107^{+21}_{-28}$ & $107^{+22}_{-23}$ \\
 Parallax (mas) & $37.647\pm0.071$ & $37.8952\pm0.00547$ & $37.8953\pm0.0060$ \\
 Barycenter $\mu_\alpha$ (mas/yr) & $-144.90^{+0.13}_{-0.13}$ & $-144.934^{+0.074}_{-0.067}$ & $-144.923^{+0.073}_{-0.070}$ \\ 
  Barycenter $\mu_\delta$ (mas/yr) & $-135.21^{+0.32}_{-0.32}$ & $-134.99^{+0.26}_{-0.24}$ & $-134.86^{+0.27}_{-0.23}$ \\ 
    RV Zero Point (m/s) & $-129^{+31}_{-68}$ & $1^{+1.6}_{-0.74}$ & $-168^{+61}_{-38}$ \\ 
    RV Jitter (m/s) & $2.53^{+0.19}_{-0.18}$ & $2.52^{+0.19}_{-0.17}$ & $17.9^{+4.3}_{-3.3}$ \\ 
\enddata
\tablecomments{Reference time for $\lambda_{\textrm{ref}}$ is 2010.0.}
\end{deluxetable*}

\begin{figure*}[t]
\centering
\includegraphics[width=0.8\textwidth]{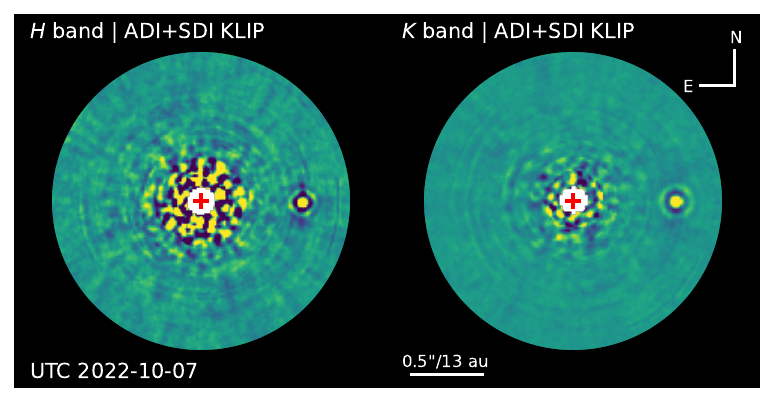}
\caption{\textbf{HD 33632 Ab imaged with Subaru/CHARIS in 2022.} Both images are post-processed with \texttt{pyklip} ADI+SDI mode and have identical color scaling. The region outside of 1.1 arcseconds has been masked for clarity. An additional source is present at the very edge of the full CHARIS field (not shown), but has been previously identified as a background star in \protect{\citet{Currie2020}}.}
\label{fig:images}
\end{figure*}

Independent measurements of the mass, age, effective temperature, and surface gravity of brown dwarfs (BDs) are exceptionally valuable for testing models of substellar formation and evolution. While temperature and surface gravity can be estimated in BDs from photometry and spectra alone (e.g. \citealt{Knapp2004,Rice2010,Martin2017_BDsurvey}), those properties cannot be translated into a precise mass determination. Furthermore, mass and age are degenerate for BDs of a given temperature and luminosity \citep{Burrows1989,Burrows2001_BDreview,Baraffe2003}, meaning it is only possible to make model independent estimates of both mass and age for BDs that occur in binaries or as companions to host stars. In systems with a host star, mass can be measured via gravitational dynamics and age can be bounded by stellar age indicators. These so-called benchmark BDs \citep{Pinfield2006_benchmarkcoining} with independent dynamical masses, plus stellar age and metallicity estimates, make a up a relatively small, but valuable fraction of the BD population. On the order of a few dozen BDs have dynamical mass measurements among thousands of identified BDs.

Directly imaged BDs that are simultaneously detected with astrometric and/or radial velocity accelerations of a host star are an important source of benchmark BDs. Astrometric acceleration and radial velocity measurements over long baselines constrain the mass of the BD, while its emission spectrum and luminosity can be directly measured with high-contrast imaging spectrometers (e.g. Gemini Planet Imager, \citealt{Macintosh2014_GPI}, VLT/SPHERE, \citealt{Beuzit2019_SPHERE}, Subaru/CHARIS, \citealt{Groff2016CHARIS}). Furthermore, long-term imaging of the BD can quantify orbital eccentricity with implications for formation history (e.g. \citealt{Forgan2018_discfrag, Bowler2020_ecc_formation}). The interest of BDs with concurrent direct imaging and spectroscopy, and measured dynamical masses, is elevated by a small, but increasing number of BDs in this category with masses or ages inconsistent with evolutionary models (e.g. Gl 229 B, \citealt{Brandt2020_Gl229,MBrandt2021}, HD 4113 C, \citealt{Cheetham2018_HD4113A}, HD 176535 Ab \citealt{Li2023_HD176535}, Lewis et al. in prep.). These objects are interesting probes of the complexity of substellar evolution since explaining them may require modifications or additions to current BD models. Unresolved binaries and factors like clouds have been suggested \citep{Li2023_HD176535} as possible explanations for the tension between models and observations.


HD 33632 Ab is a previously discovered \citep{Currie2020} BD companion to a F8V host star. The dynamical mass has been estimated using absolute astrometry from the \textit{Hipparcos-Gaia} Catalog of Accelerations (HGCA) updated for \textit{Gaia} early Data Release 3 \citep{Brandt2018_HGCA,TBrandt2021eDR3HGCA} and relative astrometry from two epochs of imaging from Keck/NIRC2 and Subaru/CHARIS published at discovery \citep{Currie2020,MBrandt2021}. A low-resolution ($R\sim19$) \textit{JHK} spectrum from Subaru/CHARIS has also been matched to observational template spectra of field BDs in \citet{Currie2020}, classifying HD 33632 Ab as a $1.7\pm0.4 \,$Gyr, $50^{+5.6}_{-5.0},\text{M}_{\text{J}}$ object near the L-T spectral transition. Concurrent with this study, high-resolution K-band spectra have been obtained with Keck/KPIC to measure the rotational and radial velocity, and chemical composition of HD 33632 Ab \citep{Hsu2024_KPICHD33632}. 

In this paper, we provide further analysis of the mass, orbit, and spectra of HD 33632 Ab utilizing a new epoch of data from the Coronagraphic High Angular Resolution Imaging Spectrograph (CHARIS), an integral field spectrograph (IFS) at the Subaru telescope, in \textit{H-} and \textit{K-} band high-resolution mode ($\sim4\times$ greater spectral resolution than the broadband mode). In Section \ref{sec:obs} we describe our new observations and the data reduction methodology, followed by relative astrometry measurements in Section \ref{sec:relast}, and orbit and mass fitting in \ref{sec:orbit}. We then extract spectra in \ref{sec:spec_ext} and fit them to templates and models in Section \ref{sec:spec_fit}, including fits to new generations of disequilibrium and cloudy BD spectral models and evolutionary tracks not explored in past work. Finally, we compare our results for the derived mass, effective temperature, and surface gravity from our analysis to previous measurements of age, and to BD evolution models in Section \ref{sec:conc}.

\section{Observations and Data Reduction} \label{sec:obs}

We observed HD 33632 A with Subaru/CHARIS \citep{Groff2016CHARIS} behind the Subaru Coronagraphic Extreme Adaptive Optics (SCExAO) system \citep{Guyon2010_SCExAO1,Vogt2010_SCExAO2} on UT 2022-10-07. A summary of the properties of the HD 33632 system are compiled in Table \ref{tab:props}. The M dwarf companion HD 33632 B is separated by $\sim34''$ ($\sim900\,$au projected distance), and is therefore not observed. We acquired 58 exposures with integration times of 40 seconds (38 minutes total) in the \textit{H} band filter in CHARIS's high-resolution mode (R$\sim75$, 20 spectral samples from $1.44746$ to $1.79091\,\mu$m) between 13:56 and 14:40 UTC, followed immediately by 69 \textit{K} band exposures (17 spectral samples from $2.01514$ to $2.36803\,\mu$m) until 15:54 UTC. The first three \textit{K} band exposures have 40-second integration times, while the remaining were increased to 60 seconds, all of which are used in data reduction ($68$ minutes total). Transit of HD 33632 occurred at 14:31 UT, such that the observations approximately bookend the time of maximum field rotation. The observations were taken with the 139 mas inner working angle (IWA) Lyot coronagraph and a 25 nm ``astrogrid'' \citep{Sahoo2020_astrogrid} added by fast deformable mirror (DM) modulation.   

Initial data reduction steps including reads-to-ramp conversion and $\chi^2$ microspectra extraction to create the 3D IFS data cubes are performed with the CHARIS pipeline described in \citet{Brandt2017CHARISpipeline}. The resulting cubes are 135 by 135 pixels with a $2''.2\times2''.2$ field-of-view ($16.15\,$ mas per lenslet). Post-processing steps including image registration using satellite spots and modelling and subtraction of the stellar PSF are performed with the CHARIS arm of the open-source \texttt{pyKLIP} package \citep{Wang2015,Chen2023}. For initial data reduction and astrometry, we utilize Karhunen Lo\'eve Image Processing (KLIP, \citealt{Soummer2012,Pueyo2016}) implementations of both angular differential imaging (ADI, \citealt{Marois2006_ADI}) and spectral differential imaging (SDI, \citealt{Racine1999_SDI}) PSF subtraction techniques. We find that signal-to-noise (S/N) of the companion is maximized around 11 KL modes and an angular and spectral movement exclusion criteria of 2 pixels. S/N is measured with the SNR map feature of \texttt{pyKLIP}, which calculates the noise in concentric annuli from the center with a small masking region to minimize bias from the real point source. Post-processed images are shown in Figure \ref{fig:images}.  

\begin{deluxetable}{llllll}[t]\label{tab:rel_ast}
\tabletypesize{\scriptsize}

\centering
\tablewidth{\columnwidth}
\tablecaption{HD 33632 Ab Measured Relative Astrometry \label{tab:BD}}
\tablehead{\colhead{Date} & \colhead{$\Delta\alpha$}& \colhead{$\Delta\delta$}& \colhead{Sep.} & \colhead{PA} \\ \colhead{(UT)} & \colhead{(mas)} & \colhead{(mas)} & \colhead{(mas)} & \colhead{($^{\circ}$)}}
\startdata
2018-10-18 & $-761\pm5$ & $-176\pm4$& $781\pm5$ & $257.0\pm0.4$\\
2018-11-01 & $-753\pm5$ & $-178\pm5$& $774\pm5$ & $256.7\pm0.4$\\
2020-08-31 & $-740\pm5$ & $-95\pm3$& $746\pm5$ & $262.8\pm0.4$\\
2022-10-07 & $-708.1\pm4.5$ & $16.4\pm5.0$& $708.3\pm4.5$ & $271.3\pm0.4$\\
\enddata
\end{deluxetable}

\begin{figure*}[t]
\centering
\includegraphics[width=\textwidth]{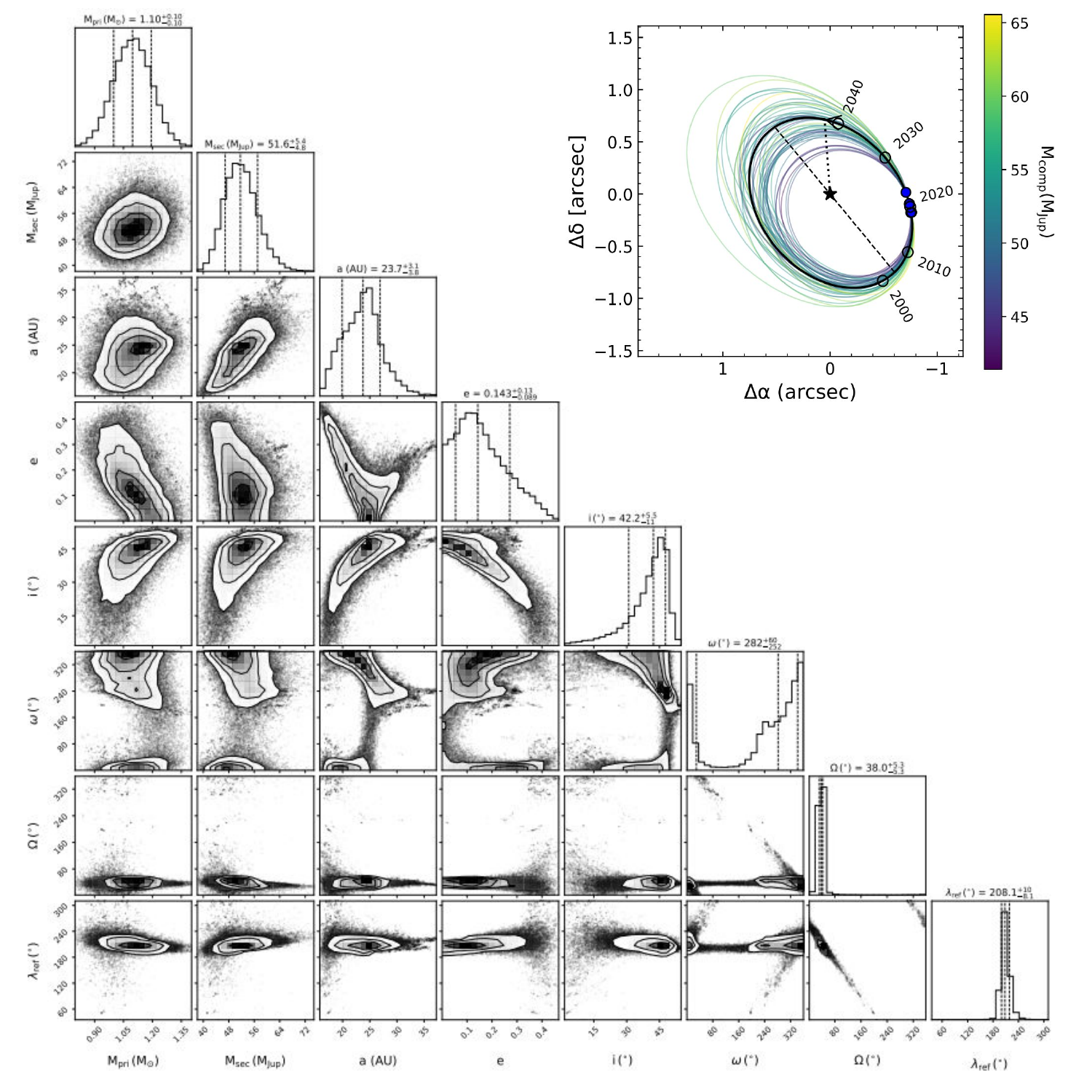}
\caption{\textbf{Posterior distributions of orbital elements and primary and secondary mass of HD 33632 Ab.} These results closely match those of \protect{\citet{MBrandt2021}}. A visualization of the orbit is shown in the upper right for varying secondary masses. The reference time for $\lambda_{\text{ref}}$ is 2010.0.}
\label{fig:corner}
\end{figure*}

\begin{figure*}[t]
\centering
\includegraphics[width=\textwidth]{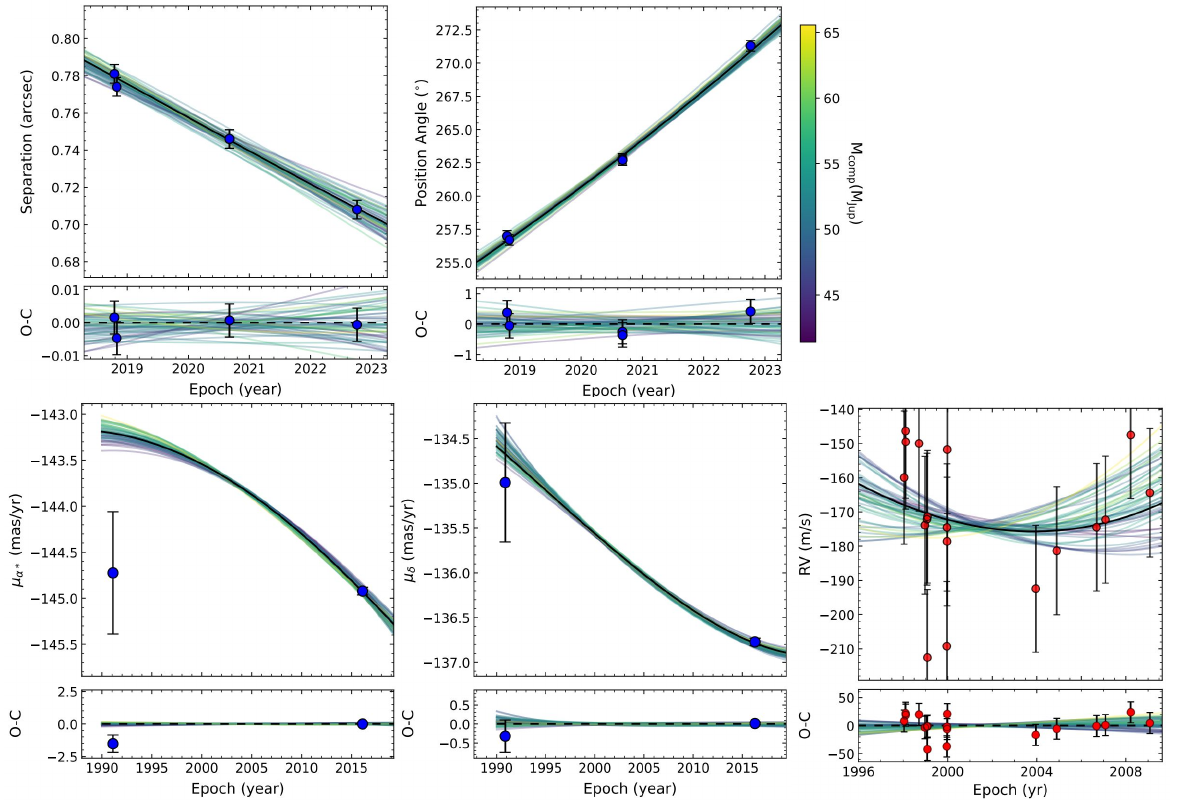}
\caption{\textbf{Derived orbits with varying secondary masses compared to astrometric and radial velocity data.} From top left to bottom right plots show separation and position angle between primary and secondary from Subaru/CHARIS and Keck/NIRC2 imaging (\protect{\citealt{Currie2020}}; this work), barycentric proper motions from \textit{Hipparcos} and \textit{Gaia} \protect{\citep{Perryman1997,GaiaDR32023}}, and radial velocities from the Lick \protect{\citep{Fischer2014}}.}
\label{fig:omc}
\end{figure*}


\section{Astrometry and Orbit Fitting} \label{sec:res}

\subsection{Relative Astrometry}\label{sec:relast}

We measure new relative astrometry of HD 33632 Ab using the Bayesian KLIP forward-model (KLIP-FM) astrometry feature of pyKLIP \citep{Wang2015}. Details of the algorithm are described in \citet{Pueyo2016} and \citet{Wang2016}, and will only be summarized here.

The KLIP-FM astrometry methodology is, at its core, PSF fitting with the critical addition of forward-modelling to account for systematic over or self-subtraction that ADI, SDI, and KLIP impart on the companion PSF which may distort the astrometry. For CHARIS, a companion PSF model is produced using the satellite spots prior to the subtraction of the stellar PSF and a best guess of the companion's spectrum. The PSF model is created as an average of all four satellite spots across all exposures on a per wavelength basis. We visually verify that the satellite spots have stable PSFs over the entire sequence. The relative brightness of the PSFs are then modulated by the approximate observed companion spectrum. The approximate companion spectrum does not need to be highly precise, with $20\%$ spectral errors causing a factor of a few less variation in the resulting astrometry than speckle noise.

This companion PSF model is used as input for Bayesian Markov chain Monte Carlo (MCMC, implemented with the \texttt{emcee} Python package, \citealt{Foreman-Mackey2013}) simulations to estimate the most likely centroid position of the companion PSF and its associated uncertainty, which can be translated to astrometric separation and position angle (PA). An additional complication of the forward-model is that stellar speckle noise (the dominant noise source) is correlated. pyKLIP considers correlated noise on the spatial extent of single speckles ($\sim \lambda / D$) using Gaussian process methods to model covariance, which is an additional fitted parameter in the Bayesian framework and is reflected in the final astrometric uncertainties.

We perform astrometry on the \textit{H}- and \textit{K}-band image cubes separately and compare the results as an assessment of bias and uncertainty. For the companion spectrum, we try a linear interpolation of the spectrum observed by \citet{Currie2020}, as well as our own retrieved spectra recursively, described in the next section, with negligible difference in the results. No astrometric calibrator (typically a well-characterized binary or compact star cluster) was observed along with HD 33632, therefore we use the results of \citet{Chen2023} for instrumental calibration of PA and platescale. We adopt the PA zero point nearest to our observations ($-2.03^{\circ}\pm0.44$), which coincidentally is the same as the average PA zero point from all tabulated measurements. \citet{Chen2023} demonstrate that the PA and platescale are stable to $0.02$ mas/lenslet and $0.08^\circ$ respectively. Our relative astrometry is presented along with previous relative astrometry in Table \ref{tab:rel_ast}.

Initial astrometry of the \textit{K}-band data resulted in an unexpected measured separation $1.5\sigma$ greater than the separation measured in \textit{H}-band astrometry, while the measured PA is well within $1\sigma$ between \textit{H} and \textit{K} image cubes. To search for a source of the discrepancy, we varied all tuneable parameters of the PSF subtraction and astrometric fit, including KLIP aggressiveness, the input spectra, contrast, and pixel position of the companion, correlated noise parameters, and MCMC chain parameters, all with inconsequential results. As a next step, we performed a leave-one-out test where we performed all post-processing and KLIP-FM astrometry steps again while sequentially leaving single data cubes out of the analysis. \textit{H}-band data proved to be exceptionally stable through this analysis with less than a single milliarcsecond variation in separation from the removal of any data cube, while \textit{K}-band data showed much greater sensitivity to individual data cubes, with as much as 5 milliarcseconds in variability ($1\sigma$ of the reported uncertainty in separation from speckle noise). We then compared the KLIP-FM astrometry to more simplistic centroid fitting with a 2D guassian to the companion PSF in the post-processed data cubes. Centroid fitting of both \textit{H}- and \textit{K}-band data gives astrometric separation consistent with the separation reported by KLIP-FM for \textit{H} band. We therefore conclude from these tests that there may be an image centering issue for the \textit{K}-band astrometry from KLIP-FM and choose to adopt the \textit{H}-band KLIP-FM relative astrometry as our data point for this epoch. We should note that adopting either the \textit{H}- or \textit{K}-band KLIP-FM astrometry from this step results in derived orbital parameters, described in the next section, that are consistent within $1\sigma$ of each other such that this issue is insignificant for final results.

We also check the relative astrometry of the background source previously identified by \citet{Currie2020}. Due to proximity to the edge of the field, we are unable to perform a KLIP-FM fit of its location, however, a simple centroid fit places it at [E,N]$''$ $= [-0.''68,-1.''32]$, only $\sim30\,$ mas from its expected location predicted solely from the proper motion of HD 33632 A. 

\subsection{Orbit Fitting and Mass Estimate}\label{sec:orbit}

We fit the orbit and mass of HD 33632 Ab using the \texttt{orvara} python package \citep{TBrandt2021ovara}, which jointly fits radial velocity (RV) observations, relative astrometry, and absolute astrometry using a Bayesian MCMC framework. This updates the work of \citet{Currie2020} and \citet{MBrandt2021} for HD 33632 with the addition of our new epoch of relative astrometry for the BD companion, but with otherwise the same methodology. This new epoch occurs approximately two years after the last epoch, covering a small, but measurable fraction of the estimated century long orbital period. We include the same data previously used by \citet{MBrandt2021} for HD 33632 Ab, namely the \textit{Hipparcos-Gaia} Catalog of Accelerations updated for \textit{Gaia} Data Release 3 \citep{TBrandt2021eDR3HGCA, GaiaDR32023}, Lick archival radial velocity data \citep{Fischer2014}, and previous relative astrometry from Subaru/CHARIS and Keck/NIRC2 observations by \citet{Currie2020}. \citet{MBrandt2021} showed that the inclusion of the outer companion HD 33632 B had insignificant impact on the orbital fit for HD 33632 Ab (and otherwise increased the complexity of the MCMC sampling); therefore, we ignore the outer companion in our orbit and mass fit. Our fitting priors are log-flat for semi-major axis, companion mass, and radial velocity jitter, gaussian for the primary mass, $\sin{i}$ for inclination, and flat for all other parameters, identical to those of \citet{MBrandt2021}. We also test uniform priors for companion mass and semi-major axis, with nearly identical results to log-flat priors.

A corner plot of the posterior distributions for the orbital elements and primary and companion masses is shown in Figure \ref{fig:corner}. Orbit plots compared to the observed relative astrometry, proper motions, and radial velocities are shown in Figure \ref{fig:omc}. The $\chi^2$ residual between the best fit orbit and observed separation and PA are $1.03$ and $3.19$ respectively. Overall, analysis with addition of our new epoch of relative astrometry produces very similar results to those from \citet{Currie2020} and \citet{MBrandt2021}. Argument of periastron, which is poorly constrained in all fits, is the only parameter not within $1\sigma$ of prior results. A summary of the derived orbital properties and companion mass is presented in Table \ref{tab:props}. Zero eccentricity cannot be ruled out, but is $<0.35$ with $95\%$ confidence.

Contemporaneous to this study, \citet{Hsu2024_KPICHD33632} have published a refined orbit and dynamical mass for HD 33632 Ab which does not include our new epoch of relative astrometry, but does consider a new direct radial velocity measurement of HD 33632 Ab. Using the \texttt{orbitize!} orbit fitting code \citep{Blunt2020_orbitize}, they report a dynamical mass of $37\pm7\,\text{M}_\text{J}$, $\sim2\sigma$ lower than our analysis. Their derived orbital parameters otherwise mostly differ by approximately $1\sigma$ from our results. We do not consider this radial velocity measurement in this paper, and a future paper should therefore re-analyze the orbit and mass using all available data. We do discuss the results of our spectral fitting in the next section in the context of both of these derived masses.

\begin{figure*}[t]
\centering
\includegraphics[width=\textwidth]{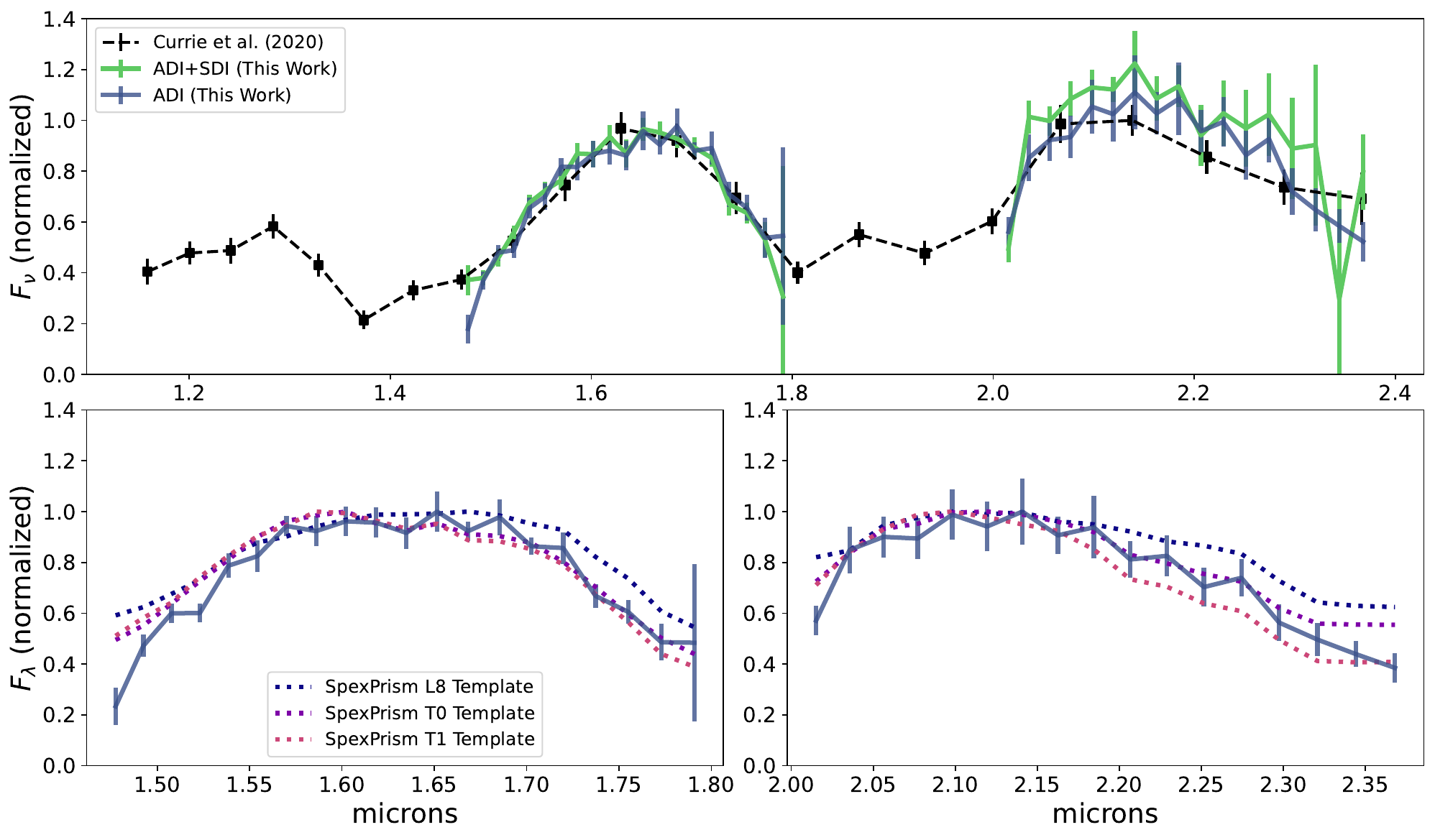}
\caption{\textbf{Top: \textit{H}- and \textit{K}-band extracted spectra from 2022 compared to the broadband spectrum from 2020 presented in \protect{\citet{Currie2020}}, also from Subaru/CHARIS.} Both datasets come from Subaru/CHARIS. \textit{H} and \textit{K} are separately scaled to best match the broadband data. \textbf{Bottom: Comparison of observed spectra (ADI only) to spectra of SpeX Prism optical and infrared standards \protect{\citep{Burgasser2017_SPLAT,sp4_Burgasser2007,sp5_Looper2007,sp2_Burgasser2004}}.} Spectral type T0 is the best fit to the data, but L8 and T1 are also plausible fits. Note the top plot is in $F_\nu$ units while the bottom plots are in $F_\lambda$ units.}
\label{fig:spexprism}
\end{figure*}

\section{Spectral Extraction and Analysis} \label{sec:diss}

\subsection{Spectral Extraction and Spectrophotometry}\label{sec:spec_ext}

\textit{H}- and \textit{K}-band spectra are extracted using the KLIP-FM functionality of \texttt{pyKLIP}. The process is similar to that used for relative astrometry, but in this case the brightness of the same wavelength dependent forward model PSF is adjusted to match the companion PSF after KLIP post-processing. More details can be found in \citet{Greenbaum2018_specFM}. We extract spectra with both the ADI+SDI and ADI-only modes to check for notable systematic differences. ADI+SDI and ADI-only spectra are consistent, although uncertainties in the ADI+SDI spectra are greater. Uncertainties on spectral data points are calculated using an injection and recovery technique. 11 copies of the forward model PSF with the same spectrum as the extracted spectrum of the real companion are injected into the data cubes before post-processing at the same separation with an even angular spacing ($30^{\circ}$ between injections). We extract the spectrum of each with the same methodology as the real companion. Our error bars are equal to the full range of flux values retrieved from the injected PSFs for each wavelength (i.e. they are more conservative than a $1\sigma$ limit of the variation). Our \textit{H} and \textit{K} extracted spectra for both ADI and ADI+SDI are shown in the top panel of Figure \ref{fig:spexprism} in comparison to the broadband \textit{JHK} spectrum from \citet{Currie2020}, also from Subaru/CHARIS.   

We perform spectrophotometry to obtain passband magnitudes for comparison to \cite{Currie2020}. The \textit{H-} and \textit{K-} band $f_{\lambda}$ spectra are multiplied by the response function of the $H_\text{MKO}$ and $K_\text{S,MKO}$ filters respectively, then integrated and converted to a magnitude using the $f_{\lambda}$ zero points \citep{Tokunaga2002_MKO2,Tokunaga2005_MKO3,Rodrigo2012_SVO,Rodrigo2020_SVO}. We measure $H_{\text{MKO}}=15.92\pm0.08$ and $K_{\text{S,MKO}}=15.54\pm0.12$. $H_{\text{MKO}}$ is within $1\sigma$ of that reported by \cite{Currie2020}, while $K_{\text{S,MKO}}$ is $\sim1.5\sigma$ fainter. The $K-$ band CHARIS spectrum does not fully cover the $K_{\text{S,MKO}}$ bandpass by a small margin, which likely explains the slight discrepancy. In any case, the discrepancy is not large enough to impact interpretation of HD 33632 Ab's spectral type.

\begin{figure*}[t]
\centering
\includegraphics[width=\textwidth]{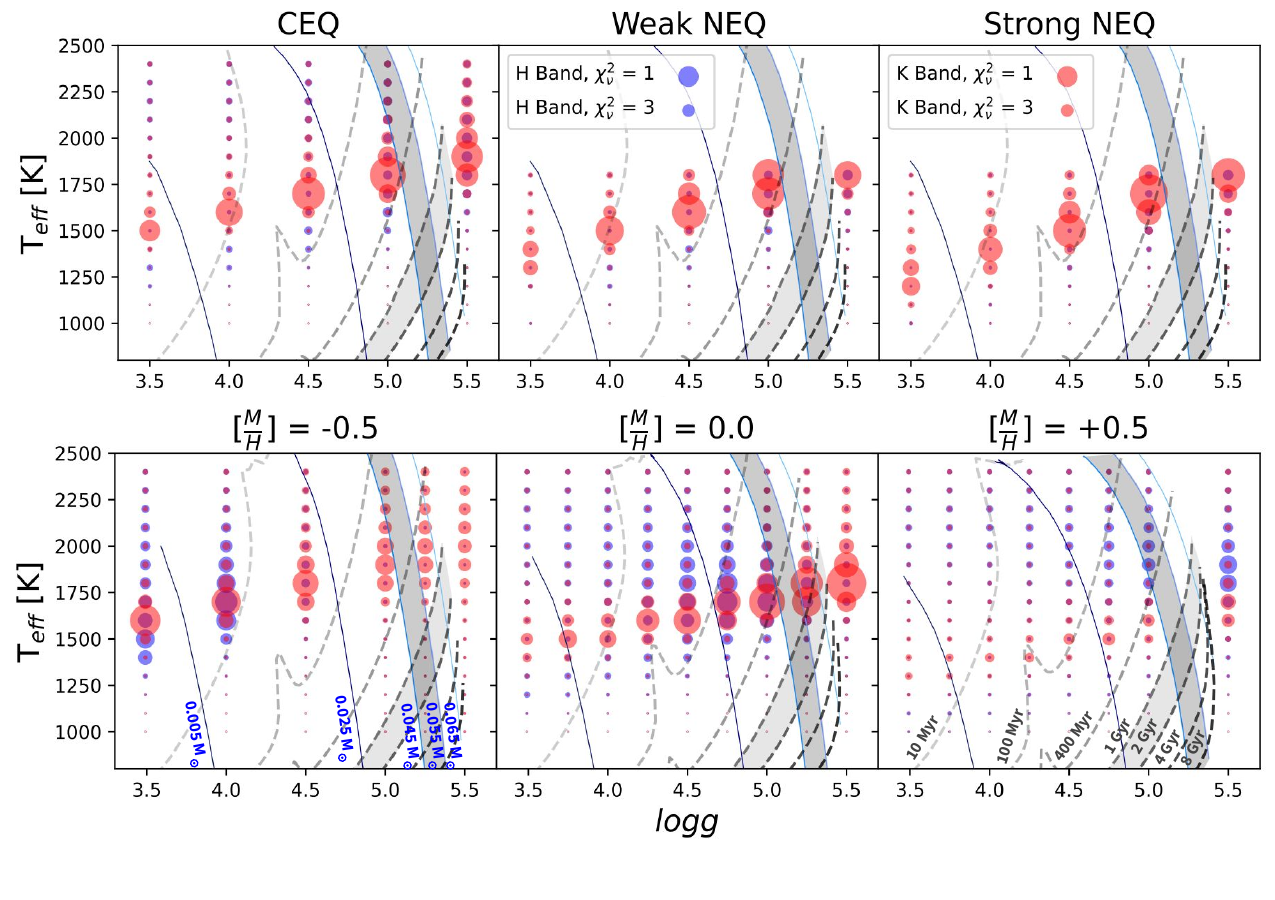}
\caption{\textbf{$\chi_\nu^2$ residuals from fits of ATMO-2020 grid models \protect{\citep{Phillips2020_atmo2020}} (top) and Sonora-Bobcat grid models \protect{\citep{Marley2021_Bobcat}} (bottom) to extracted \textit{H-} and \textit{K-}band spectral data.} \textit{H-}band residuals are blue, \textit{K-}band are red. The size of the circle is inversely proportional to the residual squared --- larger symbols indicate better fits. Columns represent models of equilibrium chemistry (CEQ), weak, and strong non-equilibrium chemistry (NEQ, vertical mixing) for ATMO-2020, and varying metallicity for Bobcat models. Contours of constant mass and age are overplotted as blue shaded and gray dotted lines respectively. Regions of the preferred mass and age from dynamical constraints and stellar age indicators are shaded, with double shading in the temperature and gravity region that agrees with both the independently-constrained mass and age. The largest circles show the preferred effective temperature and gravity from spectral fitting, which can be compared with the grey shaded regions indicating previous constraints. Fits to both ATMO-2020 and Sonora-Bobcat models prefer a hotter, younger BD than the age suggests, although they may be nearly consistent within the grid spacing. Solar metallicity Bobcat models are the best fitting and most consistent models with mass and age of the sets shown here.}
\label{fig:atmo2020}
\end{figure*}

\subsection{Spectral Fitting}\label{sec:spec_fit}

We fit the extracted spectra from \S\ref{sec:spec_ext} to BD spectral templates and atmospheric models to estimate the effective temperature, surface gravity, cloudiness, strength of disequilibrium chemistry, and C/O ratio in the atmosphere of HD 33632 Ab. When compared to evolutionary tracks in the following subsections, these variables provide a consistency check on the independent dynamical mass and age estimates for this BD, in addition to a basic characterization of atmospheric conditions. 

\subsubsection{\textit{H} and \textit{K} Spectral Fits}\label{sec:hk_fits}

To fit each template or model to the observed spectra, we arbitrarily normalize the spectral data then multiply the model by a factor such that the reduced $\chi^2$ residual ($\chi_\nu^2$) is minimized. Each set of template and model spectra come with varying native resolutions, all of which are higher than our observed spectral resolution. Each one is therefore convolved with a Gaussian and resampled to match our observed spectral resolution and sampling before it is compared to the observed data. To avoid additional spectral uncertainty from the relative flux calibration of \textit{H} and \textit{K} bands, we fit \textit{H} and \textit{K} independently, calculating a separate $\chi_\nu^2$ value for both bands for each model. 

\textit{SpeX Prism Library of Spectra} --- The SpeX Prism library is a large collection of low-resolution ($R=75-200$) observed BD and dwarf star spectra between $0.65$ and $2.55$ microns \citep{Burgasser2017_SPLAT}. We use our own code to compare our spectral data to the L and T dwarf optical and infrared standards which include 15 spectra sequentially from L0 to T8 spectral types \citep{sp1_Burgasser2006,sp2_Burgasser2004,sp3_Burgasser2007,sp4_Burgasser2007,sp5_Looper2007,sp6_Burgasser2006}. We confirm the results of \citet{Currie2020}, finding that the best fit to both \textit{H} and \textit{K} bands is the T0 template, with T1 and L8 also consistent with our spectra. A comparison of our spectra to these spectral templates is shown in the bottom panels of Figure \ref{fig:spexprism}.

\textit{ATMO-2020 Model Grids} --- ATMO-2020 is a set of cloudless 1D radiative-convective models for T and Y dwarf objects with self-consistent evolution tracks \citep{Phillips2020_atmo2020}. In addition to a grid with equilibrium chemistry, it also includes two grids with ``weak'' and ``strong'' disequilibrium chemistry that results from atmospheric vertical mixing. This disequilbrium chemistry is described by a gravity dependent $\log K_{zz}$ parameter that is greater for lower gravity objects, but is otherwise constant throughout a given atmosphere. Plots showing the $\chi_\nu^2$ parameter fit with varying surface gravity and effective temperature models are shown in the top half of Figure \ref{fig:atmo2020}. Spectral models that better fit the data (have lower $\chi_\nu^2$) are represented by larger circles, with the best fitting models having the largest circles. Fits to \textit{H-} and \textit{K-} band data are shown as blue and red circles respectively. Our \textit{K-} band spectrum is somewhat degenerate, with multiple surface gravity and effective temperature models having $\chi_\nu^2<1$, while no model has $\chi_\nu^2<1$ with our \textit{H}-band spectrum. The best fitting models are $T_\mathrm{eff}\sim1700\,$K and $\log g\sim5.0$. Overplotted in the figure are contours of mass and age from the ATMO-2020 evolutionary tracks, with the estimated mass and age limits of HD 33632 Ab shaded in grey. Based on these evolutionary models, our best spectral fits could almost be consistent with the derived dynamical mass and age for HD 33632 Ab within the grid spacing, although they are slightly ($\sim100-200\,K$) hotter than expected. Increased vertical mixing / disequilibrium chemistry slightly favors cooler models, but the results are only subtly different from equilibrium chemistry. We also fit to models with differing metallicities, the results of which are not shown for ATMO-2020, but are consistent with the metallicity results from comparison with Sonora Bobcat models described next.

\textit{Sonora Bobcat Model Grids} ---  Sonora Bobcat is another independent set of cloudless 1D radiative-convective models with self-consistent evolution, and C/O ratio as an additional model variable \citep{Marley2021_Bobcat}. We plot the result of our fits to Bobcat models in the bottom half of Figure \ref{fig:atmo2020}, which is analogous to the top half for ATMO-2020 models. While the fitting results are very similar to those of ATMO-2020, fits to Bobcat prefer a slightly cooler BD. $H-$ band data are a much better fit to Sonora-Bobcat than ATMO-2020, and prefer a BD that is $\sim50\%$ younger than stellar age estimates, although once again, it may be nearly consistent within the model grid spacing. A metal rich atmosphere produces overall much poorer fits, while a metal poor atmosphere suggests younger, lower gravity models. While there is observational evidence that HD 33632 A is metal poor \citep{Soubiran2016_PASTEL,Chen2000_pastel1,Mishenina2004_pastel2,Valenti2005_pastel3,GaiaDR32023,Gaia2023_astroparams}, the best fit models in that regime are incompatible with the estimated stellar age and dynamical mass. Only solar C/O ratios are tested for Bobcat fits, but it is a variable for fits to Sonora Elf Owl models, where low vertical mixing results with varying C/O ratios would be expected to give very similar results to that of Sonora Bobcat. High-resolution spectra from \citet{Hsu2024_KPICHD33632} support an approximately solar C/O ratio for HD 33632 Ab. 

\begin{figure*}
\centering
\includegraphics[width=\textwidth]{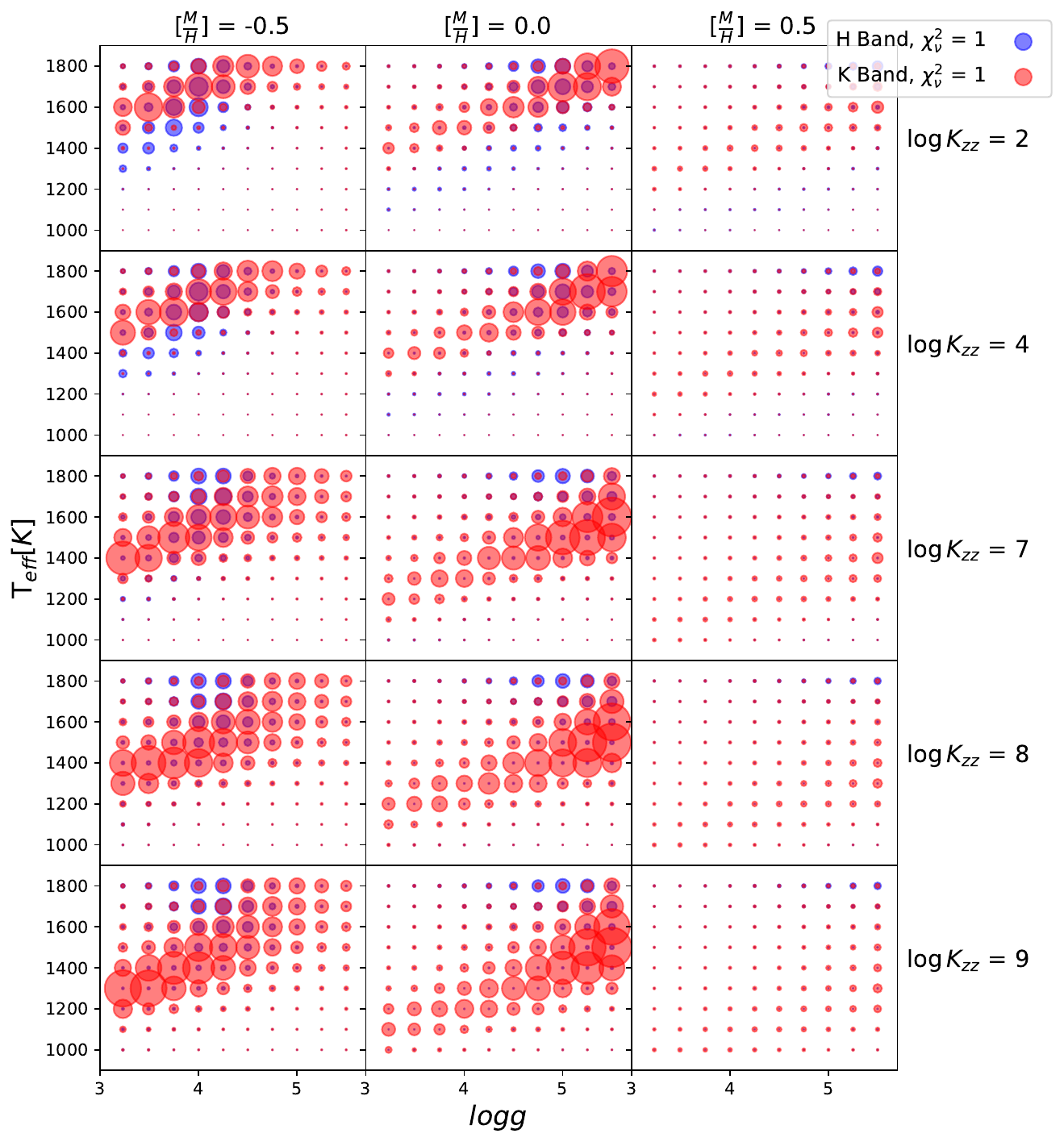}
\caption{\textbf{Same as Figure \ref{fig:atmo2020}, but for fits to Sonora Elf Owl disequilibrium models.} Increasing vertical mixing, parameterized by $K_{zz}$, prefers cooler models for fits to \textit{K} data only, while fits to \textit{H} data are less sensitive. No mass or age contours are presently available to plot for Elf Owl.  Columns represent models of differing metallicity, while rows have changing vertical mixing constants.  Plotted models have a solar C/O ratio.}
\label{fig:elfowl}
\end{figure*}

\textit{Sonora Elf Owl Model Grids} --- Sonora Elf Owl is a recent development in the Sonora family of BD model grids, which adds disequilibrium chemistry (atmospheric vertical mixing) as a variable \citep{Mukherjee2024_elfowl}. Compared to ATMO-2020 \citep{Phillips2020_atmo2020}, $\log K_{zz}$ in Elf Owl is not set by gravity and has a range of $\log K_{zz}$ for each atmosphere of a given temperature, gravity, metallicity, and C/O ratio. Our fitting results with Elf Owl are shown in Figure \ref{fig:elfowl}. The results are similar to those of Sonora Bobcat. Increased vertical mixing (greater $\log K_{zz}$) notably favors cooler models the stronger it becomes when compared to our \textit{K-}band spectrum. This is the same finding as with ATMO-2020. This could suggest that the discrepancy in age between our favored spectral models and the stellar age could be explained by increasing vertical mixing. Indeed, retrievals on high-resolution spectra of HD 33632 Ab in \citet{Hsu2024_KPICHD33632} greatly prefer strong disequilibrium chemistry. However, goodness-of-fit to our \textit{H-}band spectrum degrades with increasing $\log K_{zz}$. Therefore, we cannot claim that a high vertical mixing constant is definitively preferred to a low constant based on our data alone. We also fit against models varying in C/O ratio. Subsolar C/O results in a preference for cooler models when compared against our \textit{K}-band spectra, but these models are simultaneously poor fits to \textit{H-}band. Supersolar C/O conversely leads to a preference for hotter models, which are fit well in \textit{H} and \textit{K}, but would be less consistent with mass and age, assuming that BD evolutionary tracks for supersolar C/O are very similar to that of solar C/O. These trends are consistent across metallicity. Given this pattern, solar or subsolar C/O ratios are preferred on the basis of better agreement with dynamical mass and stellar age, which is also consistent with the finding of a roughly solar C/O for HD 33632 Ab from \citet{Hsu2024_KPICHD33632}. Note that the impact of C/O ratio, metallicity, gravity, disequilibrium chemistry, and clouds on low-resolution spectra like these can be partially degenerate, and firm constraints cannot be made without flexible forward models that can tune each parameter independently. No self-consistent evolutionary tracks are yet available to plot for Elf Owl.

\textit{Sonora Diamondback Model Grids} --- Sonora Diamondback adds clouds to the Sonora family of models, parameterized by sedimentation efficiency $f_\mathrm{sed}$ \citep{Morley2024_diamondback}. Thicker clouds have a lower value of $f_\mathrm{sed}$. Results of fitting to the Diamondback grid are shown in Figure \ref{fig:diamond}. The results clearly favor models that are either cloud-free or have thin clouds. The best models with thin clouds are slightly cooler than those of cloud-free, however, they also have slightly lower preferred surface gravities, and are therefore not necessarily better fits with the dynamical mass and stellar age. Cloud-free is a better fit to \textit{H} and \textit{K} spectra individually, but the proximity of HD 33632 Ab to the L-T transition \citep{Currie2020} indicates that it probably has at least some clouds (e.g. \citealt{Burrows2006_LT,Charnay2018_LTclouds}).

\subsubsection{Broadband \textit{JHK} Spectral Fits}\label{sec:broadband_fits}

As a check of the results from the previous subsection, we also perform the same spectral fitting analysis using the broadband spectrum ($R\sim19$) from \citep{Currie2020} that includes \textit{J-}band and will briefly summarize the results here. When the broadband spectrum is compared against Sonora Bobcat models, the fits are very poor across the grid range, with no model spectra able to replicate the relative fluxes of molecular band peaks in \textit{J, H} and \textit{K}. The relative flux in \textit{J-}band is much higher in all cloudless models than in the observed broadband spectrum, seen in Figure \ref{fig:spexprism}. The addition of clouds in the Sonora-Diamondback models reduces the relative flux of $J-$band, bringing the model spectrum into much better agreement with the data. This is strong evidence for the presence of clouds on HD 33632 Ab, which cannot be well-constrained by spectra in individual NIR bands, but has greater impact in broadband. Comparison of the broadband spectrum to Sonora Diamondback prefers an $f_\mathrm{sed}$ parameter in the range $3.0-8.0$, with thicker clouds and cloudless models being significantly worse fits. The best-fit temperature and gravity for $f_\mathrm{sed}=8.0$ is within the model grid spacing of the estimated mass and age of HD 33632 Ab, and may be fully consistent with evolutionary tracks with that cloud parameter.   

\subsubsection{Possibility of Lower Mass for HD 33632 Ab}\label{sec:lmass}

\begin{figure*}
\centering
\includegraphics[width=\textwidth]{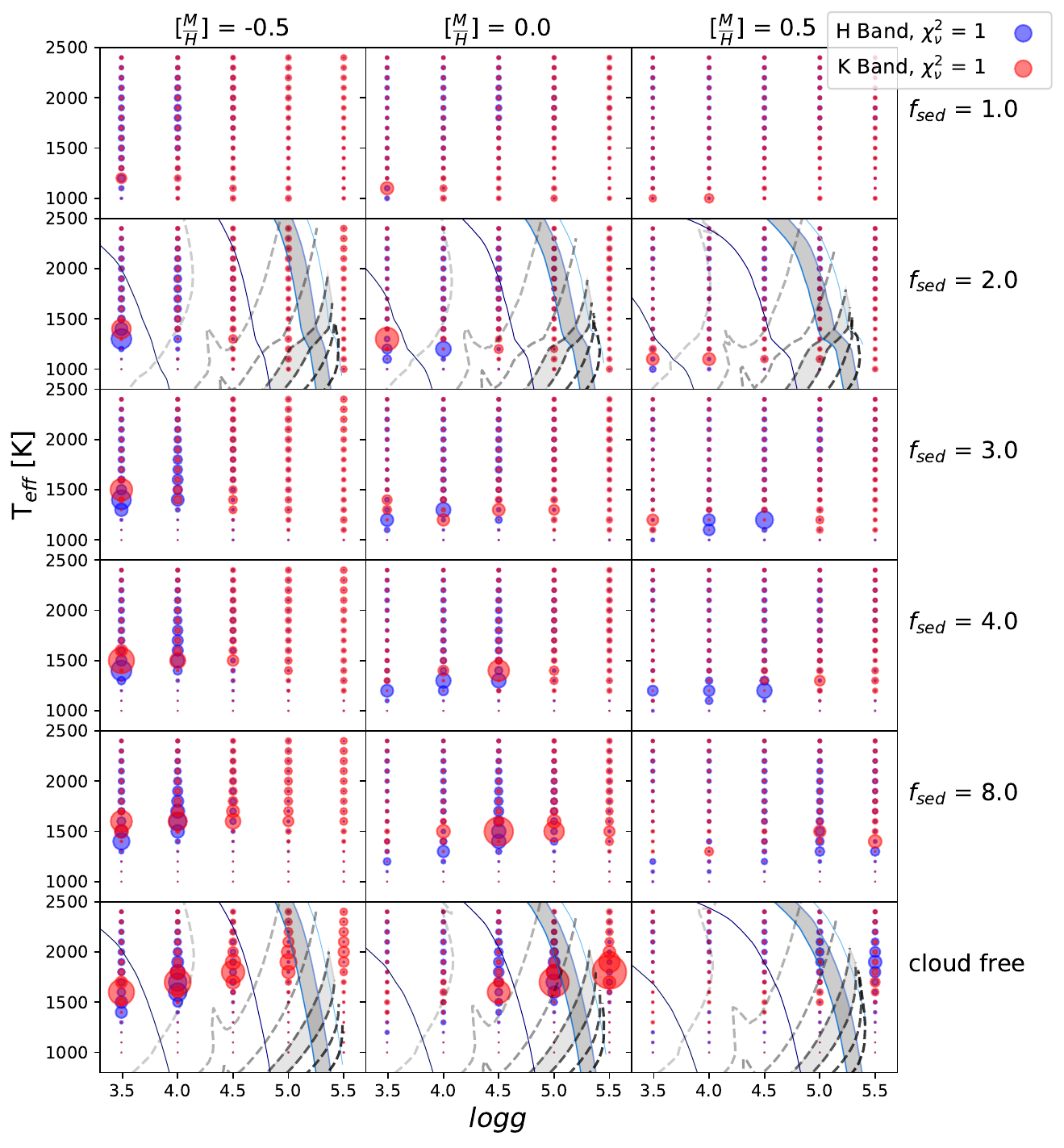}
\caption{\textbf{Same as Figure \ref{fig:atmo2020}, but for fits to Sonora Diamondback cloudy models.} Cloud free or thin clouds are preferred to the thickest clouds for fits to \textit{H} and \textit{K} data alone. As noted in Section \ref{sec:broadband_fits}, fitting Sonora Diamondback models to broadband data strongly prefers some clouds to cloud free. Lower $f_{sed}$ represents thicker clouds. Columns represent models of differing metallicity, while rows have changing $f_{sed}$, which changes cloud thickness. Plotted models have a solar C/O ratio. Only the cloudless and $f_{sed}=2.0$ models have mass and age contours supplied. The plotted contours are the same mass and age values as those plotted in Figure \ref{fig:atmo2020}.}
\label{fig:diamond}
\end{figure*}

\citet{Hsu2024_KPICHD33632} have recently published a direct measurement of the radial velocity of HD 33632 Ab, which they use to derive a $2\sigma$ lower dynamical mass than we have presented ($37\,\text{M}_\text{J}$ compared to $51.6\,\text{M}_\text{J}$). Here we briefly re-assess the best-fit spectral models in the context of the lower mass from \citet{Hsu2024_KPICHD33632}. A dynamical mass of $37\,\text{M}_\text{J}$ corresponds to a roughly $-0.25$ shift in the $log{g}$ gravity of the shaded preferred mass region of Figures \ref{fig:atmo2020} and \ref{fig:diamond}. For ATMO-2020, Sonora Bobcat, and Sonora Diamondback models, this would improve the overlap between the best-fitting models and the dynamical mass for solar metallicity and cloud-free models. There is otherwise no significant change in the preference for models with differing metallicity, C/O ratio, cloud sedimentation parameter, or strength of disequilibrium chemistry. However, this lower mass would simultaneously increase the disagreement between the age and temperature of the BD, requiring HD 33632 Ab to be either $\sim500\,$K hotter than expected for its age, or significantly younger than the estimated age of the host star. In short, a lower mass would be more in line with best-fit spectral models, but would be more difficult to explain from an evolutionary standpoint.

\section{Summary}\label{sec:conc}

We obtained a new epoch of Subaru/CHARIS imaging and spectroscopy of HD 33632 Ab at high angular resolutions \textit{H-} and \textit{K-}bands. We extract relative photometry and update the orbit and dynamical mass of the BD, finding a mass ($51.6^{+5.4}_{-4.8}\,\text{M}_\text{J}$) and orbit ($a=23.7^{+3.1}_{-3.8}\,\text{au}$ $e=0.143^{+0.13}_{-0.089}$) consistent with \citet{Currie2020} and \citet{MBrandt2021}, but a dynamical mass $\sim2\sigma$ higher than \citet{Hsu2024_KPICHD33632}. 

Spectra in \textit{H} and \textit{K} are independently compared to new generations of BD atmosphere and evolutionary models, which include disequilibrium chemistry via vertical mixing, clouds, and variable metallicity and C/O ratio. While our spectra are consistent with the broadband spectrum from \citep{Currie2020} and a spectral type near the L-T transition, no spectral model is simultaneously a best fit to the observed spectra and consistent with evolutionary tracks predicted from the derived age and preferred mass range from this study or \citet{Hsu2024_KPICHD33632}. Instead, the best spectral fits tend to prefer a $\sim50\%$ younger object than the stellar age based on a hotter derived temperature and/or lower gravity than expected. The best fit spectral models have effective temperatures of $1600-1800\,$K, whereas the stellar age ($\sim1-2.5\,$Gyr) and dynamical mass range predicts temperatures of $\sim900-1500\,$K, with lower predicted temperature for lower masses. We find that disequilibrium chemistry and light clouds independently reduce the discrepancy by favoring a lower effective temperature, though the inferred surface gravities remain slightly too low.  More complex forward modeling of the spectrum which considers both disequilibrium chemistry and clouds simultaneously, and predicts evolution, may be able to resolve the disagreement. Indeed, some plausible spectral models (i.e. Sonora Diamondback) are consistent with our derived mass and age within the grid spacing, and photometry alone is consistent with the expected luminosity of the BD from evolutionary tracks \citep{Currie2020,MBrandt2021}. We do not consider if HD 33632 Ab being an unresolved binary could be consistent with the age, mass, and spectra, however, this explanation seems presently unnecessary given the plausibility of atmospheric only explanations. 

Future imaging and data releases from \textit{Gaia} will refine the orbit and dynamical mass. Continued radial velocity monitoring should also be pursued to improve dynamical mass measurements, with precise radial velocity monitoring being more powerful in the near term than imaging alone. Foremost, our study highlights the necessity of considering complexities like clouds and disequilibrium chemistry when comparing benchmark BDs to evolutionary models, especially for BDs near the L-T transition.

\begin{acknowledgments}

This research is based in part on data collected at the Subaru Telescope, which is operated by the National Astronomical Observatory of Japan. The authors wish to recognize and acknowledge the very significant cultural role and reverence that the summit of Maunakea has always had within the indigenous Hawaiian community.  We are most fortunate to have the opportunity to conduct observations from this mountain.

This research has benefited from the SpeX Prism Spectral Libraries, maintained by Adam Burgasser at \href{https://cass.ucsd.edu/~ajb/browndwarfs/spexprism/}{https://cass.ucsd.edu/ajb/browndwarfs/spexprism/}.

This work has made use of data from the European Space Agency (ESA) mission
{\it Gaia} (\url{https://www.cosmos.esa.int/gaia}), processed by the {\it Gaia}
Data Processing and Analysis Consortium (DPAC,
\url{https://www.cosmos.esa.int/web/gaia/dpac/consortium}). Funding for the DPAC
has been provided by national institutions, in particular the institutions
participating in the {\it Gaia} Multilateral Agreement.

This material is based upon work supported by the National Science Foundation Graduate Research Fellowship under Grant No. 2021-25 DGE-2034835 for author BLL. Any opinions, findings, and conclusions or recommendations expressed in this material are those of the authors(s) and do not necessarily reflect the views of the National Science Foundation.

\end{acknowledgments}


\facilities{Subaru Telescope, National Observatory of Japan (NAOJ)}


\software{Numpy \citep{numpy}, Pandas\citep {pandas}, Scipy \citep{scipy}, Astropy \citep{astropy}}



\clearpage

\clearpage
\bibliography{references}
\bibliographystyle{aasjournal}



\end{document}